\documentclass[prl,twocolumn,letterpaper, superscriptaddress,longbibliography]{revtex4-1}
\usepackage{graphicx}
\usepackage{dcolumn}
\usepackage{bm}
\usepackage{color}
\usepackage{tabularx}
\usepackage{array}

\usepackage[utf8]{inputenc}
\usepackage[T1]{fontenc}
\usepackage{color,soul}
\usepackage{tabularx}
\usepackage{amsmath}

\makeatother
\begin{document}
\title{Hybrid Magnonics with Localized Spoof Surface Plasmon Polaritons}

\author{Yuzan Xiong}
\affiliation{Department of Physics and Astronomy, University of North Carolina at Chapel Hill, Chapel Hill, NC 27599, USA}

\author{Andrew Christy}
\affiliation{Department of Physics and Astronomy, University of North Carolina at Chapel Hill, Chapel Hill, NC 27599, USA}
\affiliation{Department of Chemistry, University of North Carolina at Chapel Hill, Chapel Hill, NC 27599, USA}

\author{Zixin Yan}
\affiliation{Department of Electrical and Computer Engineering, Northeastern University, Boston, MA 02115, USA}

\author{Amin Pishehvar}
\affiliation{Department of Electrical and Computer Engineering, Northeastern University, Boston, MA 02115, USA}

\author{Muntasir Mahdi}
\affiliation{Department of Electrical and Computer Engineering, Auburn University, Auburn, AL 36849 USA}

\author{Junming Wu}
\affiliation{Department of Physics and Astronomy, University of North Carolina at Chapel Hill, Chapel Hill, NC 27599, USA}

\author{James F. Cahoon}
\affiliation{Department of Chemistry, University of North Carolina at Chapel Hill, Chapel Hill, NC 27599, USA}

\author{Binbin Yang}
\affiliation{Department of Electrical and Computer Engineering, North Carolina A$\&$T State University, Greensboro, NC 27411, USA}

\author{Michael C. Hamilton}
\affiliation{Department of Electrical and Computer Engineering, Auburn University, Auburn, AL 36849 USA}

\author{Xufeng Zhang}
\affiliation{Department of Electrical and Computer Engineering, Northeastern University, Boston, MA 02115, USA}
\affiliation{Department of Physics, Northeastern University, Boston, MA 02115, USA}

\author{Wei Zhang}
\thanks{zhwei@unc.edu}
\affiliation{Department of Physics and Astronomy, University of North Carolina at Chapel Hill, Chapel Hill, NC 27599, USA}

\begin{abstract}

Hybrid magnonic systems have emerged as a promising direction for information propagation with preserved coherence. Due to high tunability of magnons, their interactions with microwave photons can be engineered to probe novel phenomena based on strong photon-magnon coupling. Improving the photon-magnon coupling strength can be done by tuning the structure of microwave resonators to better interact with the magnon counterpart. Planar resonators have been explored due to their potential for on-chip integration, but only common modes from stripline-based resonators have been used. Here, we present a microwave spiral resonator supporting the spoof localized surface plasmons (LSPs) and implement it to the investigation of photon-magnon coupling for hybrid magnonic applications. We showcase strong magnon-LSP photon coupling using a ferrimagnetic yttrium iron garnet sphere. We discuss the engineering capacity of the photon mode frequency and spatial field distributions of the spiral resonator via both experiment and simulation. By the localized photon mode profiles, the resulting magnetic field concentrates near the surface dielectrics, giving rise to an enhanced magnetic filling factor. 
The strong coupling and large engineering space render the spoof LSPs an interesting contender in developing novel hybrid magnonic systems and functionalities.

\end{abstract}

\flushbottom
\maketitle

\thispagestyle{empty}

\section{Introduction}

Hybrid quantum systems harness interacting excitations, such as sound waves (phonons), microwave and light waves (photons), and quantum defects (spin color centers) to complete tasks that are beyond the capability of individual systems. The collective spin excitations, i.e., spin waves (or magnons), have recently received increased attention in novel constructions of quantum hybrid systems \cite{awschalom2021quantum,yuan2022quantum,li2020hybrid,lachance2019hybrid,flebus20242024,godejohann2020magnon}. Excitations of magnetic origin, whether natural (in biomagnetism or geomagnetism) or engineered
(in magnetic devices), have long been overlooked in sensing and communication applications. But thanks to the advances in nanotechnology in the past two decades, spin excitations can be studied and manipulated down to atomic scale, leading to a range of novel technological breakthroughs. In
particular, they have recently been demonstrated as candidates for coherent information carriers and transducers even down to the quantum regime, giving rise to the emerging field of “quantum magnonics” \cite{awschalom2021quantum,yuan2022quantum,fukami2021opportunities,li2020hybrid,flebus20242024,lachance2019hybrid}. 

\begin{figure*}[htb]
 \centering
 \includegraphics[width=5.9 in]{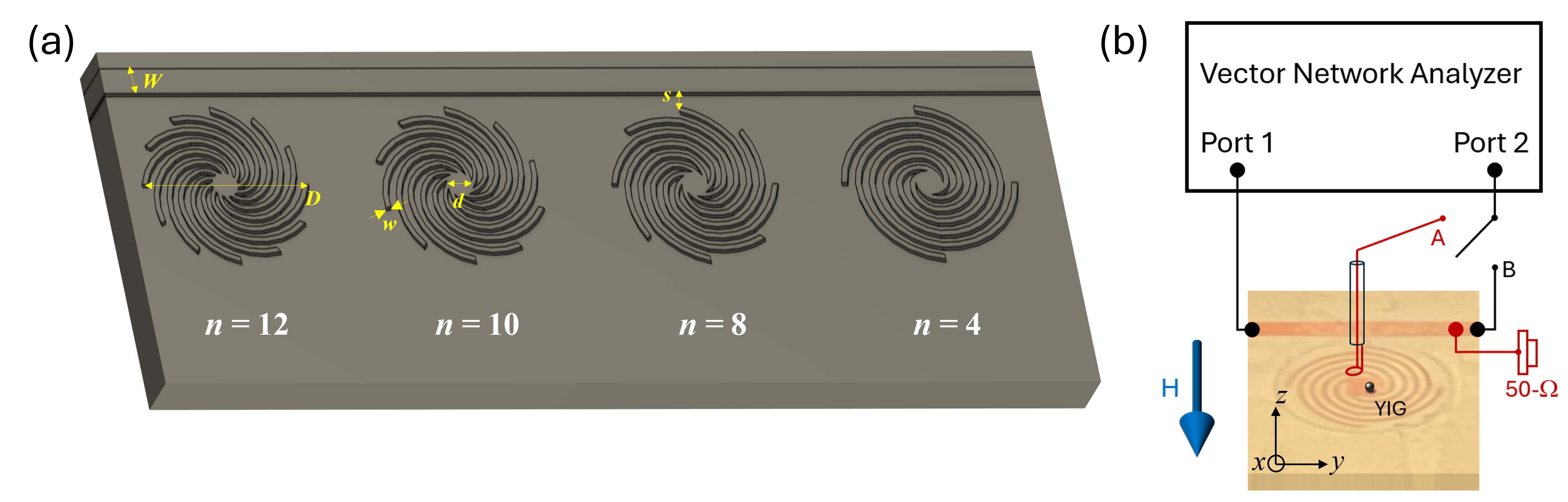}
 \caption{(a) Design patterns of the spiral photon resonators with the key dimensional parameters. $W$: excitation stripline width, $s$: separation distance between stripline and resonators, $d$: center disc aperture diameter, $D$: spiral pattern diameter, $w$: spiral arm width, and $n$: spiral arm number. (b) The sample and measurement setup: the YIG sphere is placed atop the board's surface and at different locations. One end of the resonator stripline is connected to the Port-1 of the vector-network analyzer. For global transmission measurement, the other end of the resonator stripline is connected directly to Port-2 (Switch: B). For local transmission measurement (local rf-field probing), a small probe-loop near the board surface is connected to Port-2 (Switch: A), while the other end of the stripline is 50-$\Omega$ terminated. The loop probe is anchored to a 3-D micropositioner that allows to scan along in-plane ($x,y$) and out-of-plane ($z$) directions. The in-plane magnetic field is applied perpendicular to the stripline. }
 \label{fig:design}
\end{figure*}

In the past decade, by leveraging both the engineering versatility of microwave cavities and the high tunability of magnonics, the microwave photon-magnon coupling has been serving a preeminent testbed
for exploring/demonstrating new coherent phenomena in GHz polariton sciences, including but not limited to, coherent \cite{goryachev2014high,zhang2014strongly,tabuchi2014hybridizing} and dissipative \cite{wang2020dissipative,harder2018level} couplings, nonreciprocal transmission \cite{zhang2020broadband,wang2019nonreciprocity}, remote communication \cite{rao2023meterscale,wu2021remote,li2022coherent}, magnetically-induced transparency \cite{xiong2020probing,xiong2022tunable,inman2022hybrid}, zero-reflection \cite{qian2023non}, and \textit{PT}-symmetric singularities \cite{yang2020unconventional,zhang2019experimental}. Technologically, microwave photon also benchmarks the state-of-the-art coupling strength to a superconducting qubit via cavity quantum electrodynamics (c-QED), rendering the photon-magnon systems a prominent building block for future quantum hardware construction \cite{blais2021circuit}. 

However, there is still ample space for novel engineering strategies related to photon-magnon hybridization, bestowing a large potential for further advancement in the context of quantum science applications. For example, bulky 3D microwave cavity resonators have so far been predominantly used \cite{harder2018cavity,rameshti2022cavity}. Despite their high-quality(Q) factors, their geometry inherently suffers from poor scalability. Planar resonators, on the other hand, have emerged in recent years \cite{bhoi2019photon}, driven by the increasing need for circuit-integratable systems, but only the common modes from $\lambda/2$ and lumped-element resonators were used \cite{hou2019strong,li2019strong}. In order to further benchmark the coupling efficiency and to effectively couple to unconventional excitations, tailored photon modes with matched mode profiles are of interest. Towards this end, exotic designs of photon resonators prompting unconventional mode profiles, such as surface plasmon modes \cite{polevoy2022influence}, dark modes \cite{pan2023spin}, meta-resonator mode \cite{xiong2024combinatorial,siddiky2022dual}, and spoof modes \cite{zhang2024slow}, beyond conventional geometries, have been demonstrated as promising contenders. 

The surface plasmon modes in sub-wavelength microwave resonators are electromagnetic (EM) excitations that arise from the interaction between EM waves and free electron oscillations at the interface between a metal and a dielectric material. While surface plasmon modes are more commonly associated with optical frequencies, they can also be extended to microwave frequencies \cite{zhang2020single,zhang2021spoof}. For example, by introducing periodic grooves on a metal surface, the penetration depth of the microwave can be effectively increased. This leads to the localization of the microwave signal on the surface in a way similar to surface plasmons and thus such modes are referred to spoof surface plasmon polaritons (spoof SPPs) \cite{pendry2004mimicking,Garcia-Vidal2022May}. 

Compared with conventional microwave devices such as microstrips or co-planar waveguides, the in-plane wave vectors of spoof SPPs $k_\mathrm{ip}$ are significantly increased because of the presence of the period structures. As a result, their out-of-plane wave vectors $k_\mathrm{oop}= i \sqrt{k_\mathrm{ip}^2-k_0^2}$ ($k_0$ is the free space wave vector), which are purely imaginary, have very large absolute values, corresponding to rapid decay. Although initially spoof SPPs are demonstrated for 3-D structures, it has been shown later that the same effect still applies on planar structures \cite{Shen2013Jan}. Owing to their tunability, spoof SPPs possess many exciting technological prospects, particularly in subwavelength manipulation of EM waves, such as the slow-light effect that involves stopping traveling spoof-SPPs on graded metallic structures at different positions for different frequencies \cite{gan2008ultrawide,gao2016frequency,harvey1960periodic,pendry2004mimicking}. Such a slow-wave concept has been recently implemented in hybrid magnonics at the microwave frequency utilizing the propagating spoof-SPPs modes \cite{zhang2024slow}. 

On the other hand, spoof localized surface plasmons (LSPs) can be induced by spoof plasmon resonators in the form of finite metal particles \cite{pors2012localized,shen2014ultrathin,gao2015experimental}. The LSPs are highly sensitive to the particle geometry and local dielectric environments. Therefore, their effective EM field confinement, along with an enhanced field interaction with the samples on and near the surface have rendered them especially well suited for sensing applications \cite{anker2008biosensing,liu2010planar,cai2018gain}. One of the most common realizations for the spoof-LSPs is based on the spiral geometry \cite{huidobro2014magnetic}. Spiral structures have been traditionally used as frequency-independent antennas or for polarization control at low frequencies \cite{balanis2016antenna}. Recently, spiral-based meta-resonators have also been used to improve the magnetic response of split-ring resonators \cite{baena2004artificial,bilotti2007design,xiong2024combinatorial}.

In this work, we expand the application of spoof-LSPs to hybrid magnonics and investigate magnon-photon coupling between the YIG sphere modes and the spoof-LSP modes that are supported by a spiral resonator geometry. The spiral design is intended to mimic the EM response of a corrugated metallic surface filled with grooves of a dielectric \cite{huidobro2014magnetic}. A few key geometrical parameters of the spiral resonator were varied to showcase their influence on engineering the frequency and spatial field distributions, including the center aperture size and the spiral arm number. Electric and magnetic dipole modes are observed which can strongly couple to the magnon modes of the YIG sphere. The coupling strengths were modeled and analyzed in the frame of the coupled oscillator model parameterized by the filling factor coefficient. The highly localized photon mode profiles and their large engineering space render the spoof resonators an interesting contender in developing novel hybrid magnonics systems.  

\begin{figure*}[htb]
 \centering
 \includegraphics[width=6.5 in]{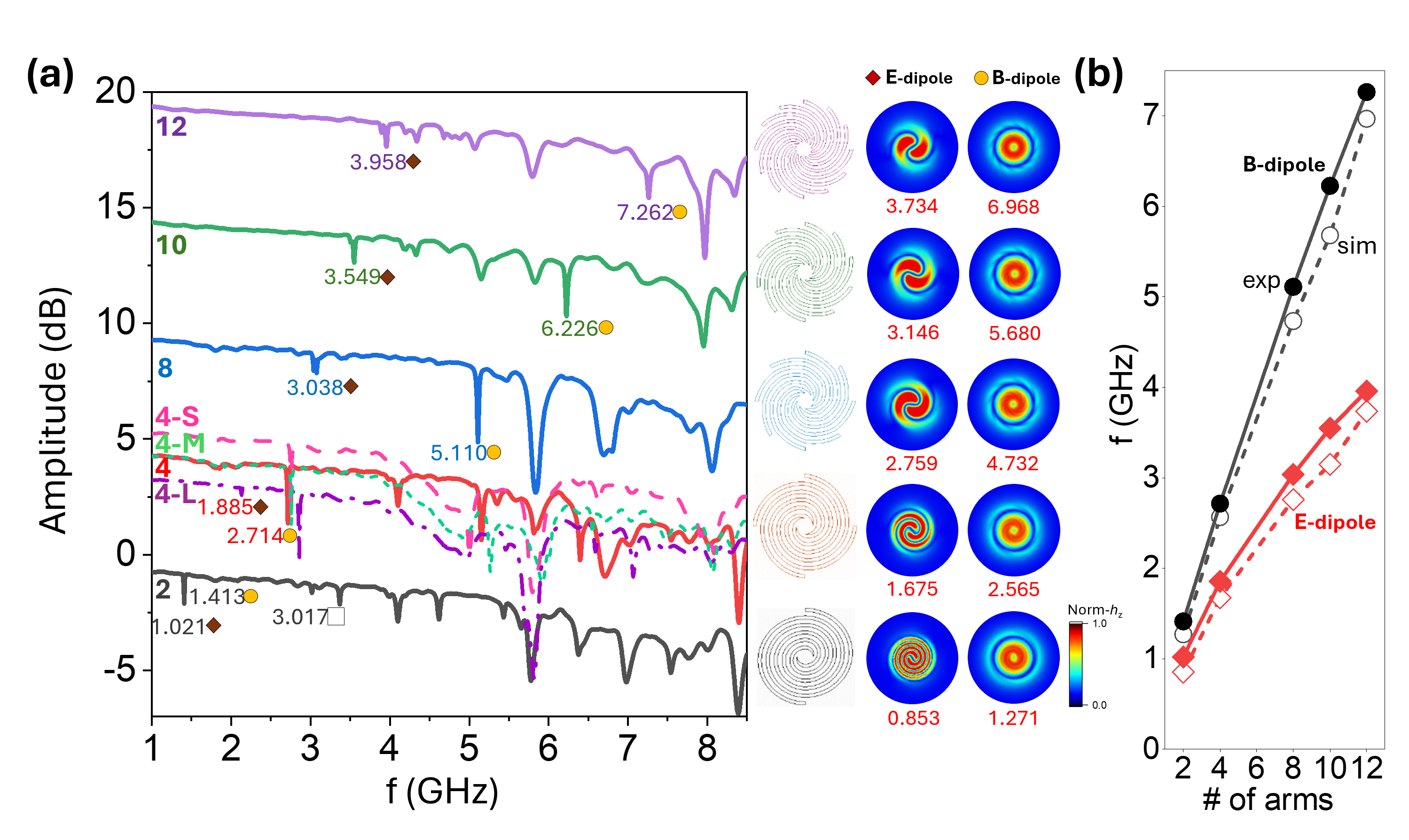}
 \caption{(a) The transmission coefficient $S_{21}$ of the spiral resonators with a center aperture size of $d$ = 1.4 mm but different arm numbers: $n =$ 2, 4, 8, 10, 12. For the 4 arm design, additional resonators with varying center aperture sizes are also shown: 4-S ($d$ = 1.0 mm), 4-M ($d$ = 2.0 mm), and 4-L ($d$ = 3.0 mm). The characteristic photon modes are labeled: E-dipole mode ($\diamondsuit$), B-dipole mode ($\bigcirc$) with the corresponding simulated $B_z$-field patterns using COMSOL (Colorbar: normalized $h_z$). (b) Comparison of the experimentally-measured (solid) and the simulated (hollow) mode frequencies for the E-dipole and B-dipole series, showing a good agreement between the experiment and simulation. }
 \label{fig:S21}
\end{figure*}

\section{Experiments}

\subsection{Spiral resonator design}

Figure 1 is the schematic representation of the spiral resonator configurations being investigated. The resonators feature a central disc aperture with a diameter ($d$) and spiral arms extending out from the center aperture. Both the aperture size ($d=1.0, 2.0, 3.0$ mm) and the number of spiral arms ($n=2, 4, 8, 10, 12$) are varied in this study. For varying aperture sizes, the arm number is fixed at $n = 4$. For varying arm numbers, the aperture size is fixed at $d = 1.4$ mm. The outer diameter ($D$) of all the resonators measures 9.0 mm. Each spiral arm is characterized by a width ($w$) of 0.3 mm, and a periodicity $p=\pi D / n$, where $n$ is the total number of arms. Such a spiral LSP geometry can be parameterized as: $x=(\frac{d}{2}+a \frac{2p}{D})\textrm{cos}(\frac{2p}{D})$ and $y=(\frac{d}{2}+a \frac{2p}{D})\textrm{sin}(\frac{2p}{D})$, where $\frac{2p}{D}$ is intersection angle of spiral line, $a$ is the intersection growth rate and $\frac{d}{2}$ corresponds to the distance between initial point of spiral line and the origin coordinates.

For efficient excitation of the spiral resonators, the separation distance ($s$) between the spiral resonators and the transmission line is 0.8 mm. The transmission line has a width ($W$) of 1.45 mm. All resonators and the microstrip that feeds them were fabricated on the Rogers TMM Laminates dielectric substrate (Rogers Corp.). The dielectric substrate has a thickness of 1.52 mm, and the metallic layer measures 35-$\mathrm{\mu}$m in thickness. The relative permittivity of the substrate is approximately 9.8. The EM properties of the spoof LSPs have been theoretically calculated using a reported analytical model \cite{liao2016homogenous}.

\subsection{Microwave Transmission Measurement}

\begin{figure*}[htb]
 \centering
 \includegraphics[width=6.5 in]{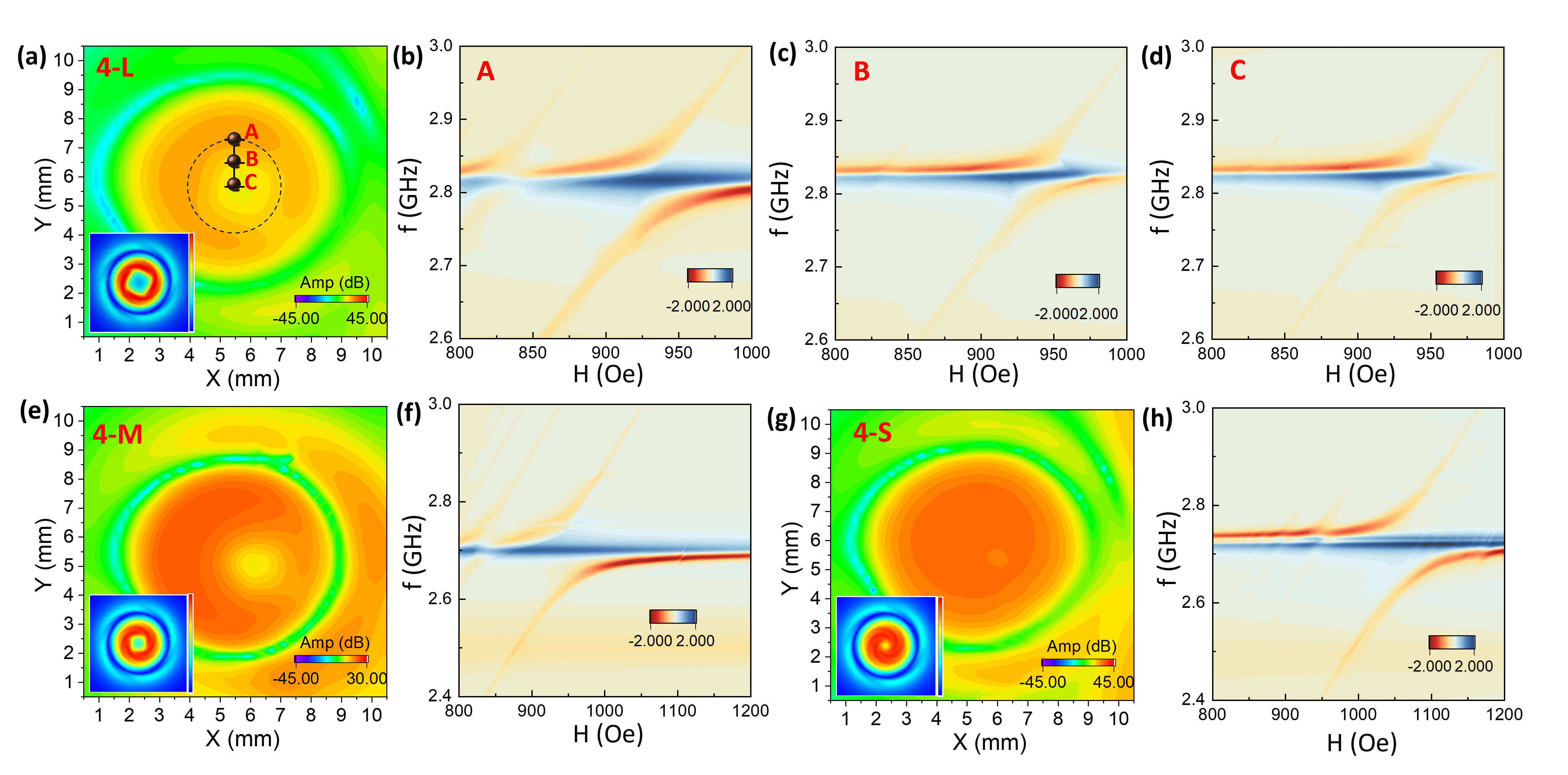}
 \caption{(a) Mapped $B_z$ field distribution at the fundamental mode using a loop probe atop the spiral resonator 4-L ($d$ = 3.0 mm, $n$ = 4). The probe is lifted $\sim 0.5$ mm above the resonator surface to obtain the near-field pattern (dashed line: the location of the center disc circumference). The YIG sphere is placed at locations: A -- C to obtain the photon-magnon coupled spectra: (b) \underline{$A$} -- along the circumference, (c) \underline{$B$} -- halfway between the edge and the center, and (d) \underline{$C$} -- at the center of the aperture. (e) Mapped $B_z$ field distribution for the 4-M resonator ($d$ = 2.0 mm, $n$ = 4). (f) Photon-magnon coupled spectra at the corresponding fundamental mode after the YIG loading. (g) Mapped $B_z$ field distribution for the 4-S resonator ($d$ = 1.0 mm, $n$ = 4). (h) Photon-magnon coupled spectra at the corresponding fundamental mode after YIG loading. The inset figures of (a), (e), and (g) show the corresponding simulated field distribution by COMSOL. }
 \label{fig:loca_disc}
\end{figure*}

\begin{figure*}[htb]
 \centering
 \includegraphics[width=6.5 in]{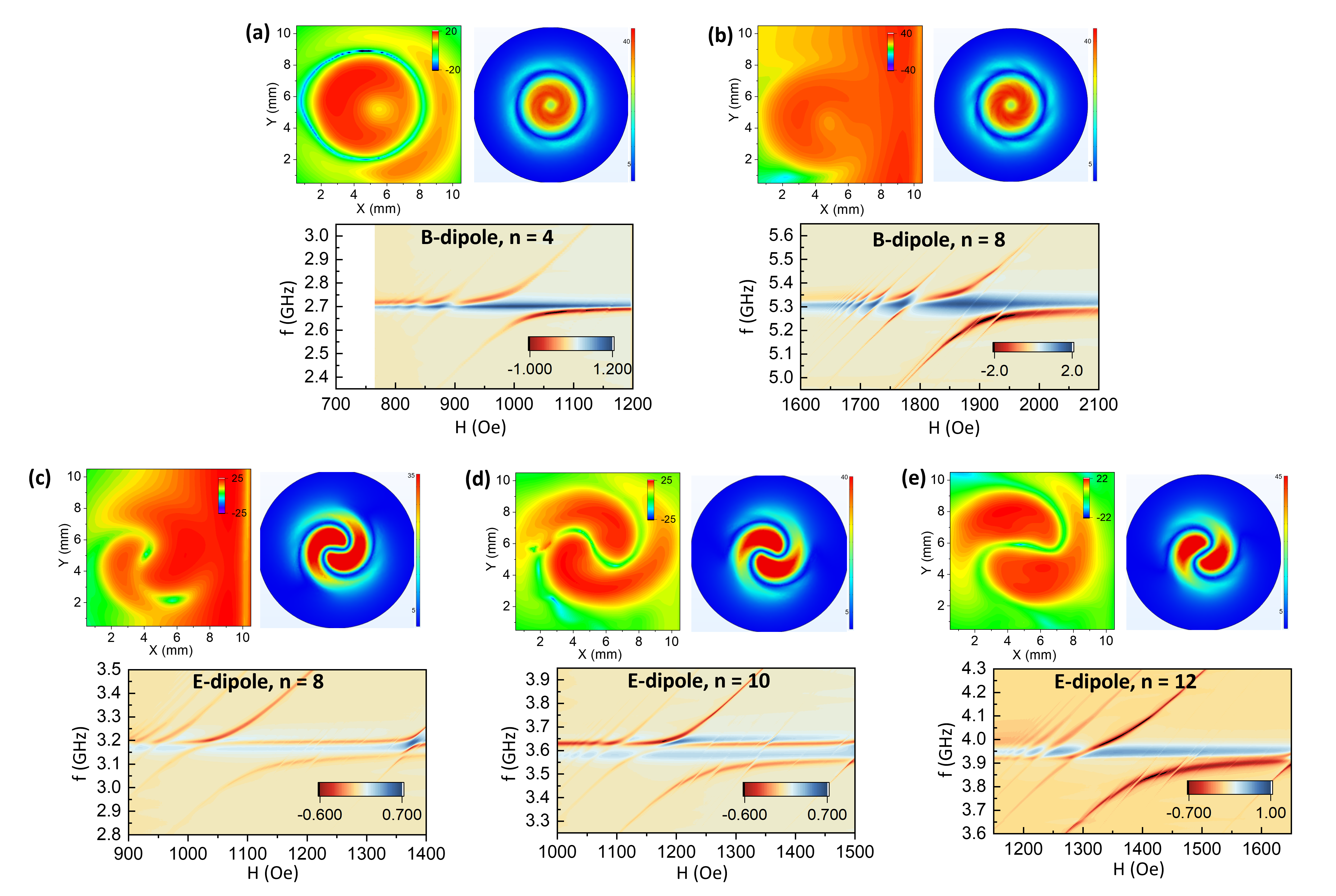}
 \caption{The measured and simulated $B_z$-field patterns of the spiral fundamental modes and the corresponding photon-magnon coupling for the varying spiral arm numbers: (a) B-dipole, $n=4$, at $\omega_\textrm{4B}$ = 2.71 GHz, (b) B-dipole, $n=8$, at $\omega_\textrm{8B}$ = 5.32 GHz, (c) E-dipole, $n=8$, at $\omega_\textrm{8E}$ = 3.10 GHz, (c) E-dipole, $n=10$, at $\omega_\textrm{10E}$ = 3.57 GHz, (e) E-dipole, $n=12$, at $\omega_\textrm{12E}$ = 3.92 GHz. }
 \label{fig:finger}
\end{figure*}

Microwave properties of these resonators without magnetic field measurements were characterized by using a vector-network analyzer (VNA). The measured transmission coefficient $S_{21}$ of all the spiral resonators are summarized in Fig. \ref{fig:S21}(a). The resonances of the spiral resonators show up in the transmission spectra as Lorentzian or Fano-shaped transmission dips. To identify different resonance modes, finite-element method (FEM) simulations are performed using the eigenmode solvers of the RF module in COMSOL. The spiral resonators are modeled based on the actual dimensions of the fabricated devices, and the PCB substrate is modeled as a dielectric slab with a dielectric constant of 10. To identify the resonance frequencies and mode profiles, the metallic structures are modeled as perfect electric conductors (PECs) with the Ohmic loss ignored. Sufficiently large simulation regions and scattering boundary conditions are used to suppress undesired boundary reflections. Meshes are refined near the spiral structures to guarantee good accuracy and resolution. Each simulation gives multiple eigen-frequencies, corresponding to different modes of the spiral resonator. In our simulation of the spiral resonators, the YIG sphere is ignored due to its relatively small size which only causes perturbation to the field distribution. Accordingly, only Maxwell’s equations are used in the simulations, while the Landau-Lifshitz-Gilbert equation that governs the spin wave (magnon) dynamics is not included. For each eigen-frequency, all the EM field components are obtained from the built-in Maxwell’s equations in the RF module and can be directly plotted in COMSOL, including the $B_z$ field, allowing the direct observation of the mode profile for each resonance.

\begin{table}[htb]
\centering
\begin{tabular}{|p{1.0cm}|p{1.0cm}|p{1.0cm}|p{1.0cm}|p{1.0cm}|p{1.0cm}|p{1.0cm}|}
\hline
     & \multicolumn{2}{c|}{Fundamental} & \multicolumn{2}{c|}{2nd-order} & \multicolumn{2}{c|}{3rd-order} \\\hline
arm \# & E-\tiny dipole & B-\tiny dipole & E-\tiny dipole & B-\tiny dipole & E-\tiny dipole & B-\tiny dipole \\\hline \hline
2  & 1.021 & 1.413 &  \textbf{3.017}  & 3.368 & 4.621  & 5.792 \\\hline
4  & 1.885 & \textbf{2.714} &  4.572  &       &        &       \\\hline
4-S  & 1.816 & \textbf{2.751} &         &       &        &       \\\hline
4-M  & 1.917 & \textbf{2.757} &         &       &        &       \\\hline
4-L  & 2.130 & \textbf{2.858} &         &       &        &       \\\hline
8  & \textbf{3.038} & \textbf{5.110} &         &       &        &       \\\hline
10 & \textbf{3.549} & 6.226 &         &       &        &       \\\hline
12 & \textbf{3.958} & 7.262 &         &       &        &       \\\hline
\end{tabular}
    \caption{The experimentally identified E- and B-dipole LSP modes for the resonators ($n = $2, 4, 8, 10, 12). The unit is in GHz. The bold numbers are modes used for subsequent photon-magnon coupling investigations. Other modes are either too low (outside the YIG magnon Kittel dispersion) or too high (outside the VNA testing band). Only dipolar modes are listed here. Additional quadruple modes will be discussed in Appendix A. } 
    \label{tab:all_modes}
\end{table}

For each resonator, a series of dipolar resonance modes, either electric(E) or magnetic(B) type, can be identified from the plot and their frequencies are detailed in Table 1. Additional quadruple modes are also observed for $n \ge 4$, which will be discussed in Appendix A. For both the E- and B-dipole modes, the frequency increases with the number of arms according to both the experiment and simulation, Fig. \ref{fig:S21}(b). For the present study, we focus on the fundamental E- and B-dipole modes. For photon-magnon coupling studies, we use a YIG sphere with a nominal diameter of 1.0 mm, and also apply an external magnetic field, $H$, perpendicular to the transmission line, to satisfy the ferromagnetic resonance condition; however, for modes with a frequency lower than 2.2 GHz (as for the $n =$ 2 and 4 designs), the YIG magnon modes do not overlap with the photon modes. 

In addition, to elucidate the magnetic($B$)-field distribution of the LSP modes, we map out the out-of-plane $B$-field component ($B_z$) using a loop probe custom made from a coaxial cable assembly with an unterminated end. The loop diameter is estimated $\sim$ 0.8 mm. The measurement setup and $B_z$-field maps at the mode frequencies of interest are detailed in Appendix A and B.    

\subsection{Photon-Magnon Coupling}

Given the peculiarities of the LSPs mode structures, the spatial field distribution of such resonators is an important parameter to their coupling to magnetic samples. For such structures, the resonantly induced rf current from each spiral arm overall converges on the circumference of the inner disc aperture, where the magnetic component of the EM field is the strongest. This should in principle lead to a pronounced radial dependence.  

To verify this important trait, we study the resonator with the largest aperture size ($d = 3.0$ mm) and use 4 spiral arms as an example. We mapped out the $B_z$ field distribution at the fundamental B-dipole mode using a loop probe, near the surface of the board and directly above ($\sim 0.5$ mm) the spiral resonator, Fig. \ref{fig:loca_disc}(a). The $B_z$ amplitude exhibits a quasi-uniform distribution in the shape of concentric circles with an amplitude rapidly decaying from the circumference to the center of the aperture, consistent with the characteristic of a B-dipole mode \cite{liao2016homogenous}. Then, we placed the YIG sphere at different representative locations and measured the photon-magnon coupling. A series of Walker modes can be prompted which couple to the spoof-LSP mode. In such photon-magnon coupled system, a strong coupling is realized if the coupling strength, $g$, exceeds the dissipation of both the photon ($\kappa_p$) and the magnon ($\kappa_m$) counterparts, manifesting an anti-crossing feature in the observed spectrum. \cite{li2020hybrid}  

We first concentrate on the Kittel magnon mode. The coupling strengths are found by analyzing the anticrossing gap (formed between the LSP mode and the Kittel mode), yielding: \underline{$A$} -- along the circumference, $g_A$ = 50.2 MHz, Fig. \ref{fig:loca_disc}(b), \underline{$B$} -- halfway between the edge and the center, $g_B$ = 26.9 MHz, Fig. \ref{fig:loca_disc}(c), and \underline{$C$} -- at the center of the aperture, $g_C$ = 26.8 MHz, Fig. \ref{fig:loca_disc}(d). The coupling strength rapidly decays as the YIG is moved away from the disc edge. The higher-order Walker modes share a similar trend to the Kittel mode, and will be addressed in more detail in the Discussion and Appendix C. 

In addition, by placing the YIG sphere at different locations along the circumference, we can study the angular dependence of the LSP-magnon coupling. In the present case (B-dipole), we did not observe any notable variation in the coupling strength, which is consistent with the uniform angular distribution of the $B_z$ field (indicated by both the experimental field mapping and the simulation). For the E-dipole mode to be discussed later, the $B_z$ field exhibits a quasi-dipolar distribution, and the coupling strength is attenuated by $\sim 30 \%$ at the minimum position (short-axis) compared to the maximum (long-axis). Thus, we use the maximum aperture edge location for all of our subsequent investigations, which is the position that benchmarks the coupling strength for our herein designed resonators.

Next, the results for the additional 4-M and 4-S resonators are shown to illustrate the effect of the center aperture size, see Fig. \ref{fig:loca_disc}(e-h). As mentioned earlier, the generic resonance frequencies before the YIG sphere loading are found at $\omega_\textrm{4-S}$ = 2.751 GHz, $\omega_\textrm{4-M}$ = 2.757 GHz, and $\omega_\textrm{4-L}$ = 2.858 GHz, respectively. Such values agree reasonably well with the COMSOL simulations, at 2.571, 2.592, and 2.651 GHz. For spoof-LSP resonators with fixed radii, the arm length reduces as the disc diameter increases, accordingly leading to increased resonance frequencies. The systematic deviation of the simulated resonance frequencies relative to the measurement results (by roughly 200 MHz) can be attributed to the geometrical size variation of the actual fabricated devices from the models used in the design and simulation. The mapped $B_z$-field patterns for 4-M, Fig.\ref{fig:loca_disc}(e), and 4-S, Fig.\ref{fig:loca_disc}(g), indicate the same quasi-uniform distribution with concentric circles.

After the YIG sphere loading, the resonance frequencies slightly shifted, with $\omega_\textrm{4-S}$ = 2.72, $\omega_\textrm{4-M}$ = 2.71, $\omega_\textrm{4-L}$ = 2.83 GHz, respectively. The coupling strengths between such LSP mode and the YIG's Kittel mode are found by analyzing the anticrossing gap, yielding: Fig.\ref{fig:loca_disc}(f),$g_\textrm{4-M}$ = 74.8 MHz, and (h) $g_\textrm{4-S}$ = 77.2 MHz, both larger than the previous $g_\textrm{4-L} = g_A$ = 50.2 MHz, in (b). The dependence of coupling strength on the disc size can be explained by the mode profile of the LSPs obtained from COMSOL simulations. The microwave magnetic fields are more concentrated near the disc on devices with smaller discs, consequently enhancing their coupling with the magnon modes in the YIG sphere that is located on or near the disc.

The most prominent dependence was found in varying the number of the spiral arms: the results for five different arm numbers, 2, 4, 8, 10, and 12 were compared in Fig. \ref{fig:finger}. All resonators have a center disc size of 1.4 mm. For the $n=4$ resonator design, the fundamental E-dipole mode is too low to couple with the YIG magnons, so we focus on the fundamental B-dipole mode. For the $n=8$, 10, 12 resonator designs, we focus on the E-dipole modes coupling with magnons. Both the fundamental E- and B-dipole modes couple strongly with the YIG's Kittel modes with very similar magnitudes, as evidenced by the $n=8$ resonator measurement.   

For all the rest arm numbers, after the YIG loading, the resonance frequencies slightly shifted, see Fig.\ref{fig:finger}. The coupling strengths between those respective LSP modes and the YIG's Kittel mode are found by analyzing the anticrossing gap, Fig.\ref{fig:finger}, yielding: (a) $g_\textrm{4B}$ = 47.2 MHz, (b) $g_\textrm{8B}$ = 82.4 MHz, (c) $g_\textrm{8E}$ = 84.6, (d) $g_\textrm{10E}$ = 99.6, and (e) $g_\textrm{12E}$ = 115.5 MHz, respectively. 

Evidently the coupling strength is enhanced as the number of spiral arms increases, which can be attributed to the enhanced mode localization. At the center, the microwave fields from different arms constructively interfere, forming the localized mode. When the number of arms increases, the spacing among neighboring arms decreases, leading to enhanced inter-arm coupling which enhances the mode intensity near the center disc, and accordingly the magnon-spoof-LSP coupling.

\begin{figure*}[htb]
 \centering
 \includegraphics[width=7.0 in]{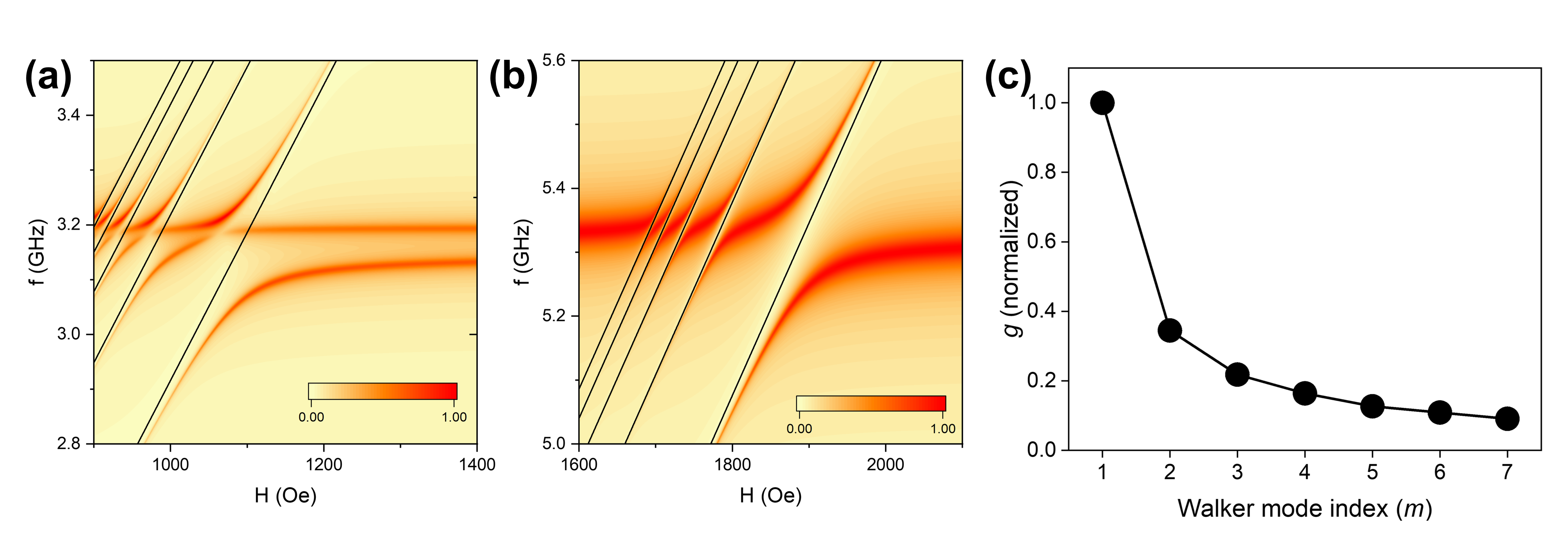}
 \caption{The calculated photon-magnon coupling using the coupled harmonic oscillator model for (a) E-dipole, $n$ = 8, (b) B-dipole, $n$ = 8. Walker modes up through $m$ = 5 were used with $M_0$ = 1690 Oe, $H_A$ = 480 Oe, and $\gamma$ = 27.0 GHz/T, determined from the experimental data. The solid black lines are the Walker modes used for the calculations. (c) The relative coupling strengths used in the model for the $n$ = 8 resonator extracted from the experimental data. Coupling strengths for couplings up to $m$ = 7 were able to be resolved.}
 \label{fig:model}
\end{figure*}

\section{Discussions}

In the context of photon-magnon coupling, an important figure-of-merit is the magnetic filling factor ($\xi$), which represents the amount of overlap between the microwave photon magnetic field and the collective spin profile in the YIG sphere sample. The coupling strength $g$ is directly related to the filling factor, where $g = \omega_i \sqrt{ \chi_{eff} \xi_i}$, with $\omega$ being the resonator eigen-frequency and $\chi_{eff}$ the effective magnetic susceptibility of the YIG \cite{goryachev2014high}. Therefore, qualitative arguments can be made through the discussion of the effective filling factor of the pertinent LSP modes, under the varying experimental parameters outlined above (center disk size, YIG sphere placement, number of spiral arms, etc.).

Table \ref{tab:my_label} summarizes the fitted eigen-mode frequencies (with YIG sphere loading), coupling strengths, and the derived filling factor ratio ($\xi_n / \xi_0$), in correspondence to the different varying experimental parameters. 

\begin{table}[h!]
    \centering
    \begin{tabular}{|p{2cm} p{2cm} p{2cm} p{1cm}|} 
        \hline
        \textbf{\# of Arms (n)} & \textbf{Eigen-mode (GHz)} & \textbf{Coupling (MHz)} & \textbf{$\xi_n / \xi_0$} \\
        \hline
        2 & 1.41 (B) & N/A & N/A \\
        \hline
        4 & 2.71 (B) & 47.2 & 1.00 \\
        \hline
        8 & 3.08 (E) & 84.6 & 2.59 \\
        \hline
        10 & 3.56 (E) & 99.6 & 2.56 \\
        \hline
        12 & 3.96 (E) & 115.5 & 2.85 \\
        \hline
        \textbf{YIG Sphere Placement} &  &  &  \\
        \hline
        Center & 2.83 (B) & 26.9 & 1.00 \\
        \hline
        Halfway & 2.83 (B) & 27.0 & 1.01 \\
        \hline 
        Edge & 2.82 (B) & 50.2 & 3.52 \\
        \hline
        \textbf{Center Disk Size} &  &  &  \\
        \hline
        Large & 2.83 (B) & 47.9 &  1.00 \\
        \hline
        Medium & 2.69 (B) & 74.8 & 2.86 \\
        \hline
        Small & 2.74 (B) & 77.1 & 2.95 \\
        \hline
        \end{tabular}
    \caption{Comparison of fundamental frequencies, coupling strengths, and filling factor ratios of resonators based on explored experimental parameters. $\xi_0$ is the baseline value ($n$ = 4, edge YIG placement, large center disk) used for each section of the table. For $n$ = 2, the demonstrated photon-magnon interaction occurs with the 2$^\textrm{nd}$ order B-dipole mode.}
    \label{tab:my_label}
\end{table}

To further elucidate the involved photon and magnon modes and their interactions, we developed an analytical model using the `coupled-oscillator approach' that well reproduces all the experimentally obtained photon-magnon spectra. In this framework, each involved photon and magnon mode is represented as its own damped harmonic oscillator. The oscillators are bound together by springs with coupling strength \textit{g}, which is proportional to the anti-crossing gap sizes from the experimental data. The photon modes have field-independent frequencies, so they appear horizontal in the experimental and modeled spectra. The magnon modes have a field dependence, so they are represented using the Kittel equation and Walker mode theory. The finer details of the model can be found in Appendix C.

For devices with 8, 10, and 12 spiral arms, the E-dipole shows coupling to the YIG Kittel mode. The azimuthal order $n_z$ = 2 causes a mode splitting, reflected in Fig. \ref{fig:finger}(c-e), where the major photon modes interacting with the YIG Kittel mode split in two, denoted as $\omega_+$ and $\omega_-$  \cite{gao2016frequency}. Both photon modes can couple to the YIG Kittel and higher-order Walker modes. The multiple couplings can be modeled with the described `coupled harmonic oscillator' approach, where each photon and magnon mode is represented as its own oscillator. The $n \times n$ coupling matrix can be written in the form of:  

\begin{equation} \label{eq:discmatrix}
\begin{pmatrix}
\omega - \tilde{\omega_1} & g_{12} & ... & g_{1n} \\ g_{12} & \omega - \tilde{\omega_2} & ... & g_{2n} \\ ... & ... & ... & ... \\ g_{1n} & g_{2n} & ... & \omega - \tilde{\omega_n}
\end{pmatrix}
\end{equation}

As an example, our modeling of the $n$ = 8 E- and B-dipole couplings are discussed here, as shown in Fig. \ref{fig:model}(a-b). The rest of the modeling can be found in Appendix C. Due to the localized mode profiles (also reflected in our measurements and simulations), the magnetic field strength and distribution are non-uniform near the YIG sphere, which excites a set of magnon modes beyond the standard Kittel mode \cite{morris2017strong}. These modes manifest as the (\textit{m}, \textit{m}, 0) Walker modes, whose frequencies can be represented as \cite{fletcher1959ferrimagnetic}: 

\begin{equation}
\omega = \gamma [H - H_A + 4\pi M_0 (\frac{m}{2m + 1})]
\end{equation}
where $H_A$ represents the total contribution from anisotropy and $M_0$ is the saturation magnetization. The magnetization value is derived from the spacing of the modes in the experimental data. The effect of anisotropy is a constant shift of the Walker mode frequencies, and its magnitude was determined by fitting Eq. 2 to each set of magnon dispersion. The photon-magnon coupling strengths are all greater than either the photon or magnon linewdiths for up to $m = 7$ of the Walker modes, i.e., in the strong coupling regime. Their values (relative to the Kittel mode coupling strength) are shown in Fig. \ref{fig:model}(c). For the purpose of computational simplicity, this model analyzes the Walker modes up to $m = 5$ as shown in Fig. \ref{fig:model}(b). The coupling strengths monotonically decrease as the mode number $n$ increases. A similar trend has been observed in planar stripline resonators coupling to a YIG sphere \cite{morris2017strong}. 

Finally, for the model of the E-dipole to show the same behavior at the resonance point between the split photon modes and the YIG Kittel mode, the coupling strengths of $\omega_+$ and $\omega_-$ to the Kittel mode must not be equal. This is also reflected in Fig. \ref{fig:finger}(c), where a non-uniform magnon mode that couples to the split modes at 1370 Oe has different coupling strengths to each of them. We attribute this to different levels of overlap between their mode profiles and that of the magnon modes. This same logic applies for each higher order Walker mode, with successive decreases in the overall coupling strengths, due to less overlap of both $\omega_+$ and $\omega_-$ with the higher \textit{m} modes. 

In summary, we investigate magnon-photon coupling between the YIG sphere modes and the spoof-LSP modes. The LSPs are supported by a winding spiral resonator geometry, whose key design parameters were varied to showcase their influence on engineering the frequency and spatial field distributions, including the center aperture size and the spiral arm number. Electric and magnetic dipole modes are observed which strongly couple to the magnetic Walker modes of the YIG sphere. The coupling strengths were modeled and analyzed in the frame of the coupled oscillator model parameterized by the filling factor coefficient. The highly localized photon mode profiles and their large engineering space render the spoof resonators an interesting contender in developing novel hybrid magnonics systems. Although the present demonstration only focuses on the lower GHz range, the design principle can potentially extend to higher frequencies, such as in the sub-THz regime where couplings to antiferromagnetic magnon modes are of interest. \cite{han2023coherent,bialek2021strong,bialek2023cavity,khatri2023220}    

\section{Acknowledgements}

Experimental work at UNC-CH was supported by Air Force Office of Scientific Research (AFOSR) under award number FA2386-21-1-4091 and the U.S. National Science Foundation (NSF) under Grant No. ECCS-2246254. X.Z. acknowledges support from NSF (2337713) and ONR Young Investigator Program (N00014-23-1-2144).

\section{Appendices}

\subsection{A. Magnetic field mapping and simulation}

\renewcommand{\theequation}{A-\arabic{equation}}
\setcounter{equation}{0}  
\renewcommand{\thefigure}{A-\arabic{figure}}
\setcounter{figure}{0}  

In the main text, we mainly addressed the fundamental E- and B-dipole modes and their photon-magnon couplings. Higher-order E- or B-modes were also identified in the transmission and spatial mapping experiments, especially for the larger arm numbers. Here, as a general characterization, we append the results of the comprehensive magnetic field mapping for each resonator. 

\begin{figure}[htb]
 \centering
 \includegraphics[width=3.5 in]{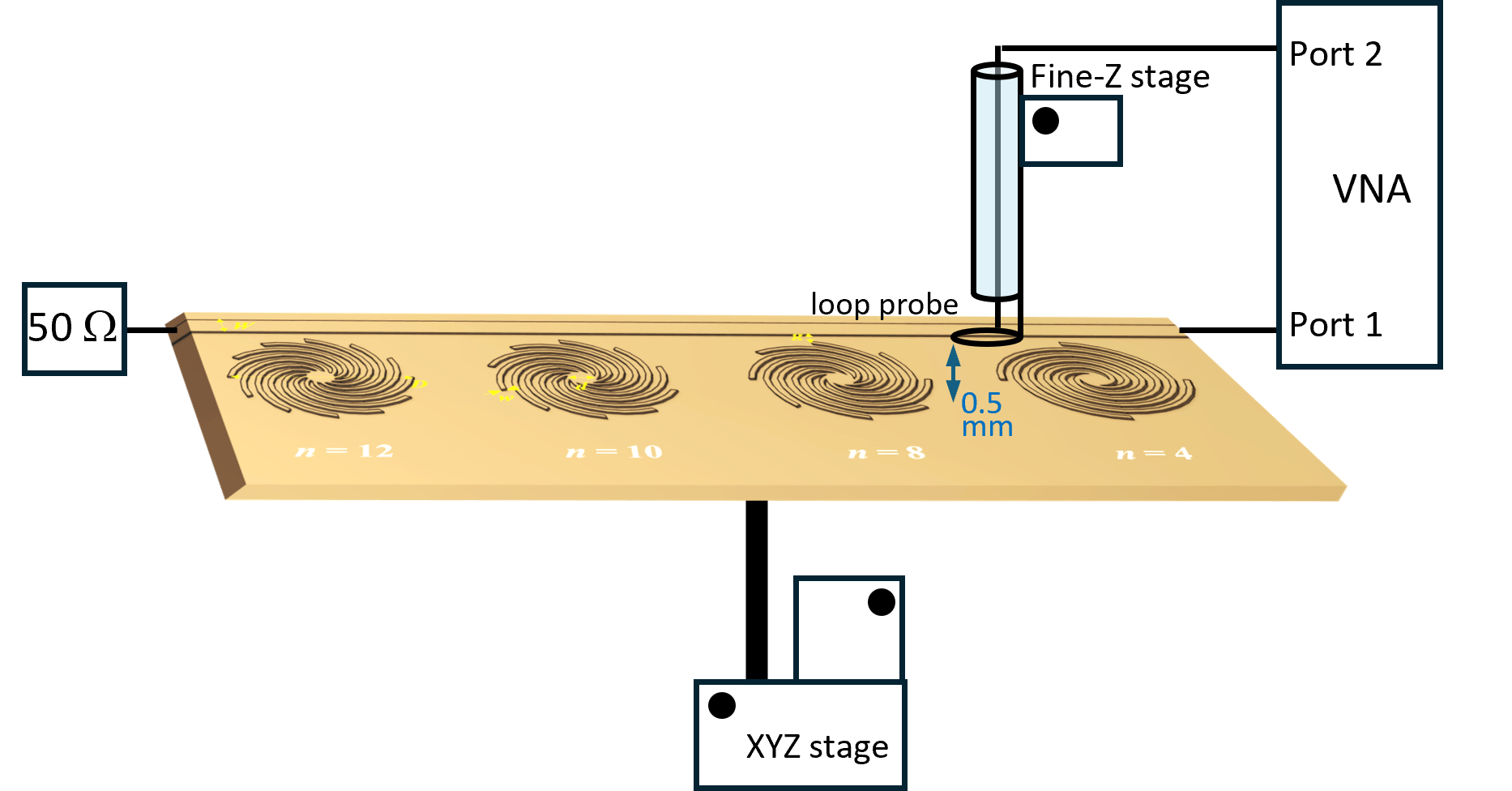}
 \caption{Schematic of the experimental setup for mapping out the out-of-plane magnetic field ($B_z$) distribution. }
 \label{fig:map_setup}
\end{figure}

Figure \ref{fig:map_setup} illustrates the detailed setup for mapping the $B_z$-field distribution. The microwave signal inputs from Port 1 of the VNA to the stripline and terminates at the other end using a 50-Ohm cap. The Port 2 is connected to a microwave loop probe anchored to a precise $z$-stage. The initial height is set $\sim 0.5$ mm atop the board surface. The position of the board is controlled by another $x$-$y$ stage that anchors the board.    

\begin{figure}[htb]
 \centering
 \includegraphics[width=3.6 in]{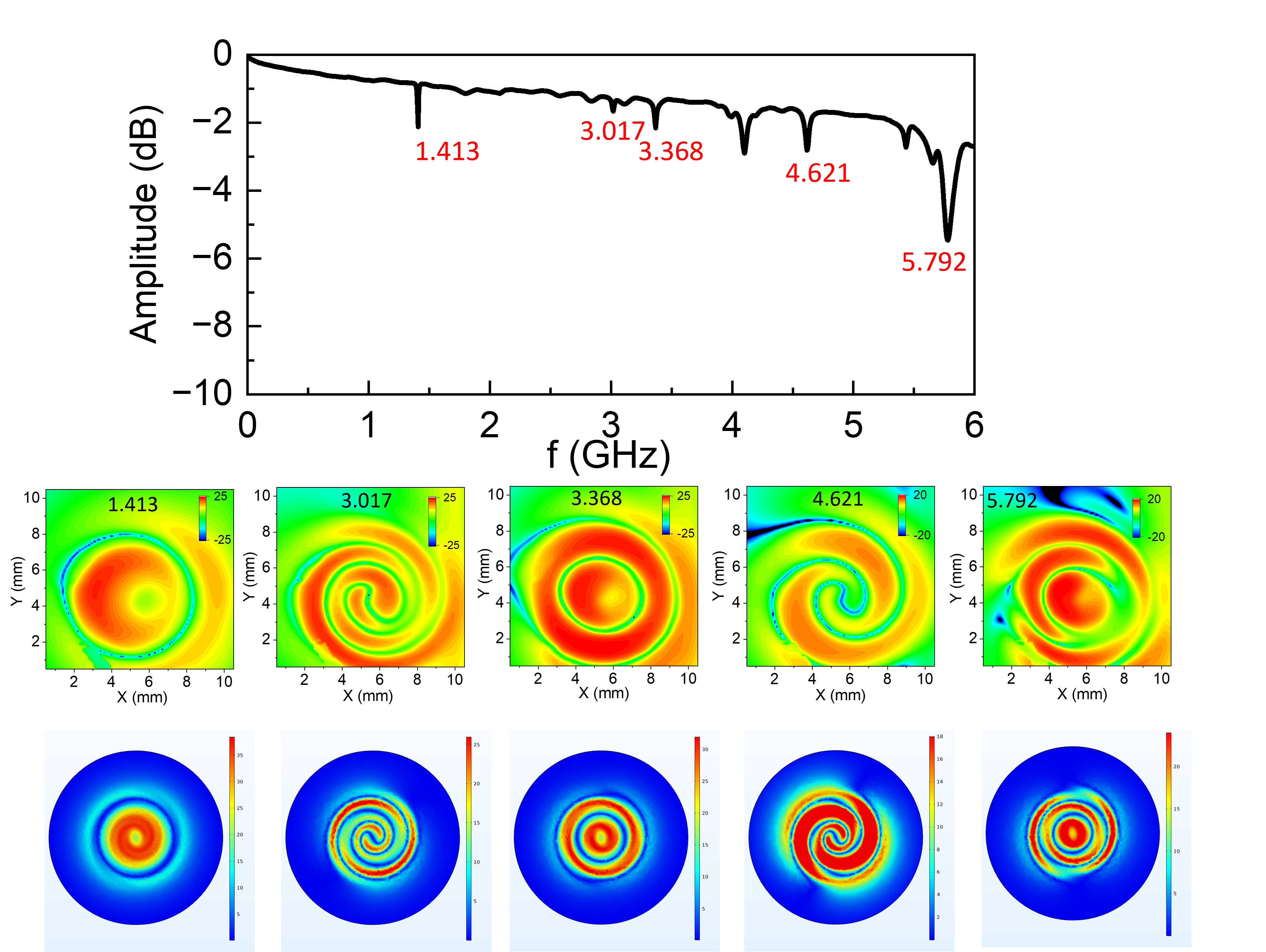}
 \caption{The transmission characteristic $S_{21}$ (top) and the measured $B_z$-field maps and the corresponding simulated field distributions at selective resonance frequencies (bottom) for the $n=2$ resonator.}
 \label{fig:2}
\end{figure}

Figure \ref{fig:2} shows the measured $B_z$-field maps and the corresponding simulated field distributions at selective resonance frequencies for the $n=2$ resonator. In particular, the 2nd- and 3rd-order E- and B-dipole modes can be clearly identified from their spatial field distribution: the 2nd and 3rd E-dipole: at 3.017 and 4.621 GHz, respectively, featuring the double winding spiral geometry, and the 2nd and 3rd B-dipole: at 3.368 and 5.792 GHz, respectively, featuring concentric ring structures. 

\begin{figure}[htb]
 \centering
 \includegraphics[width=3.6 in]{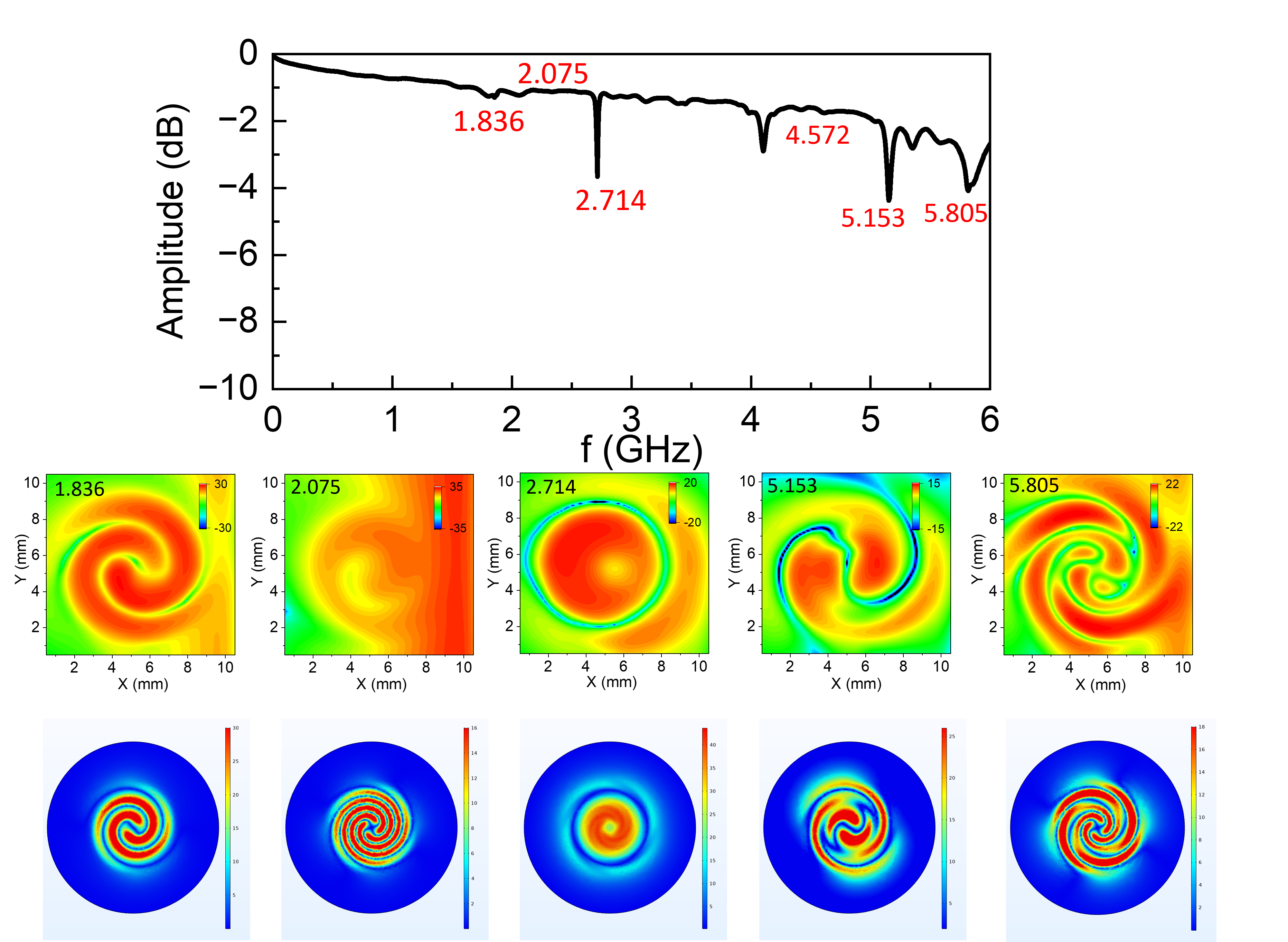}
 \caption{The transmission characteristic $S_{21}$ (top) and the measured $B_z$-field maps and the corresponding simulated field distributions at selective resonance frequencies (bottom) for the $n=4$ resonator.}
 \label{fig:4}
\end{figure}

Figure \ref{fig:4} shows the measured $B_z$-field maps and the corresponding simulated field distributions at selective resonance frequencies for the $n=4$ resonator. The map at 2.075 GHz exemplifies an intermediate state as the mode evolves from a double winding (E-type) to the concentric ring (B-type) structure. In addition, the fundamental quadruple mode is identified at 5.805 GHz, which cannot be supported by the $n=2$ design. 

\begin{figure}[htb]
 \centering
 \includegraphics[width=3.6 in]{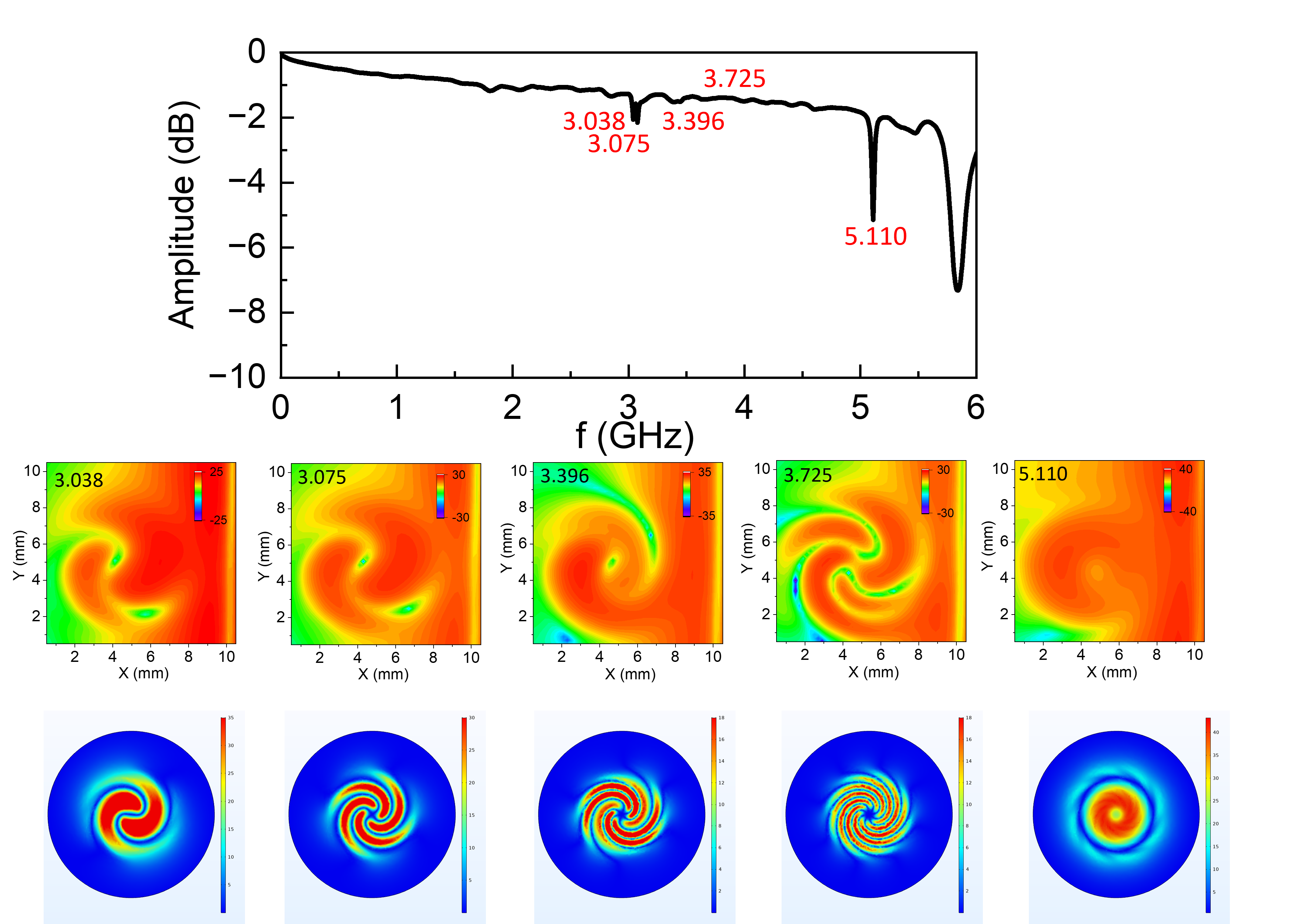}
 \caption{The transmission characteristic $S_{21}$ (top) and the measured $B_z$-field maps and the corresponding simulated field distributions at selective resonance frequencies (bottom) for the $n=8$ resonator.}
 \label{fig:8}
\end{figure}

\begin{figure}[htb]
 \centering
 \includegraphics[width=3.4 in]{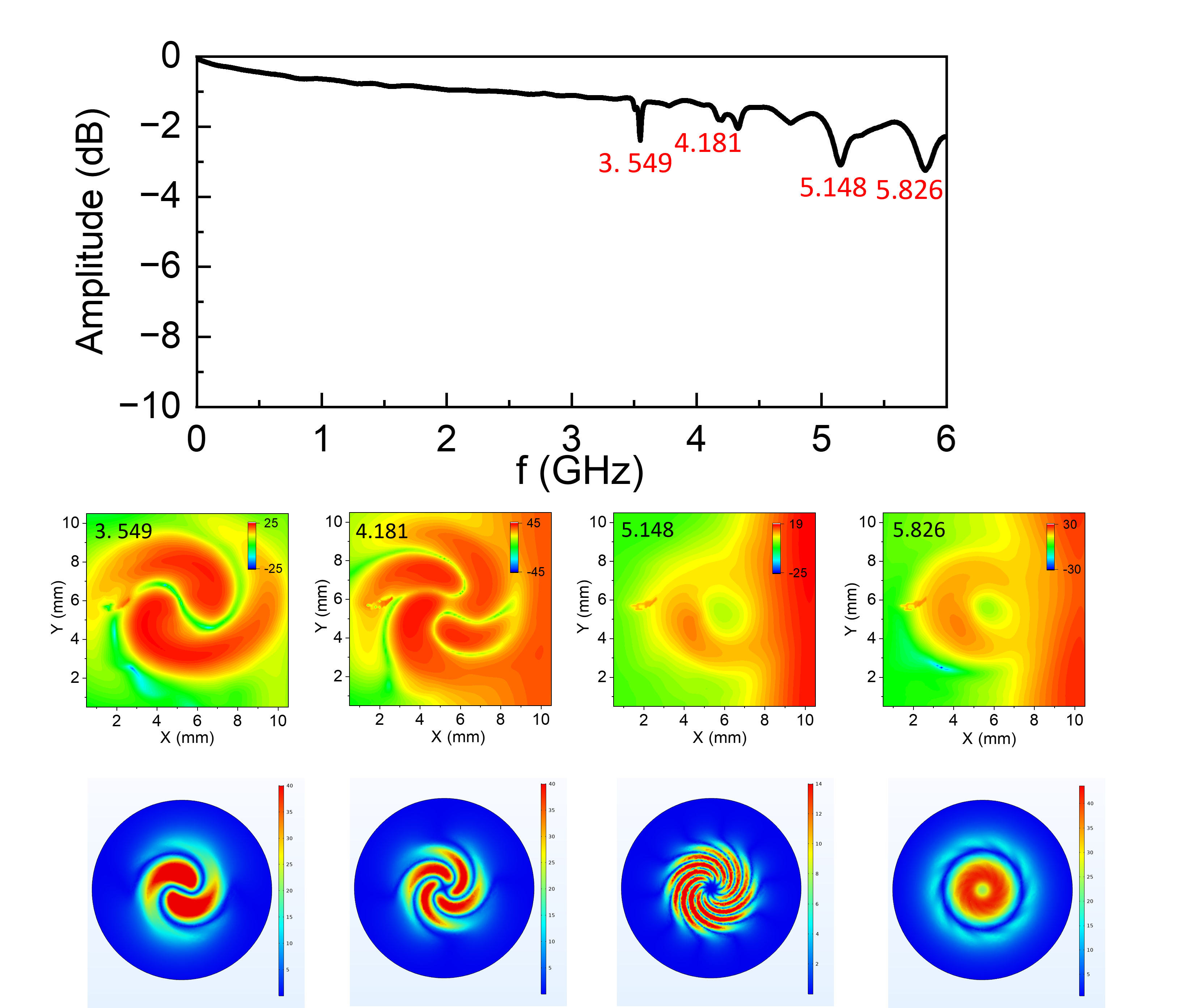}
 \caption{The transmission characteristic $S_{21}$ (top) and the measured $B_z$-field maps and the corresponding simulated field distributions at selective resonance frequencies (bottom) for the $n=10$ resonator.}
 \label{fig:10}
\end{figure}

\begin{figure}[htb]
 \centering
 \includegraphics[width=3.6 in]{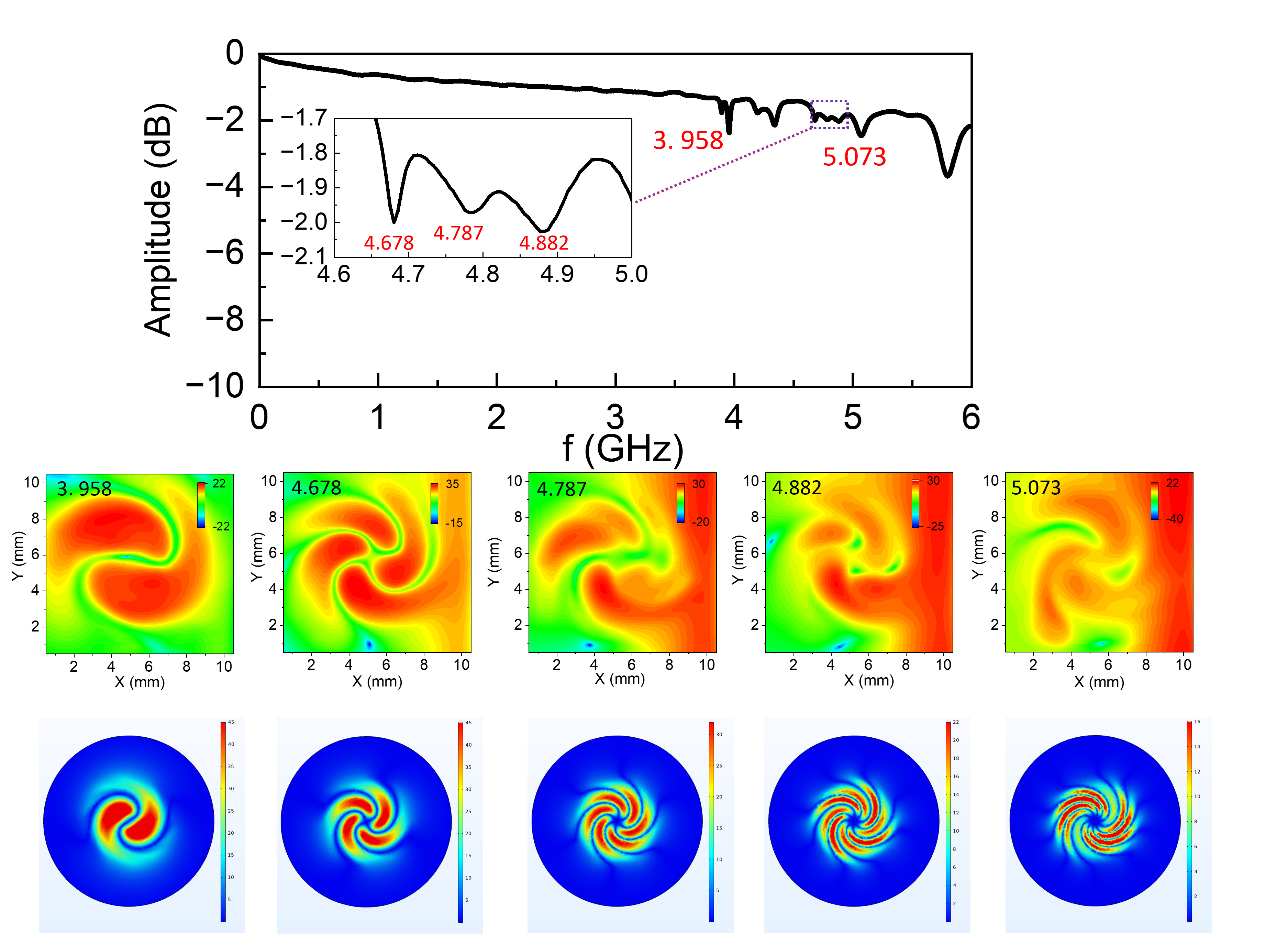}
 \caption{The transmission characteristic $S_{21}$ (top) and the measured $B_z$-field maps and the corresponding simulated field distributions at selective resonance frequencies (bottom) for the $n=12$ resonator.}
 \label{fig:12}
\end{figure}

Figure \ref{fig:8} shows the measured $B_z$-field maps and the corresponding simulated field distributions at selective resonance frequencies for the $n=8$ resonator. A few representative frequencies, at 3.075, 3.396, and 3.725 GHz, were picked out to show the evolution from the fundamental E-dipole to the B-dipole mode. In addition, double peaks near the fundamental E-dipole modes (3.038 and 3.075) were observed, which present for $n=8$ and $n=12$ resonators but not for $n = 2, 4, 10$.

Figure \ref{fig:10} and \ref{fig:12} show the measured $B_z$-field maps and the corresponding simulated field distributions at selective resonance frequencies for the $n=10$ and $n=12$ resonators, respectively. More complex field map patterns at the intermediate frequencies are observed as the modes evolve from the E- to the B-dipole fundamental modes. For the $n=12$ resonator, the intermediate mode at 4.678 has a clear quadruple shape, but an adjacent mode at 4.787 indicates likely a hexapole character, which was also confirmed by the COMSOL simulation. Such a hexapole geometry is uniquely supported by this arm number. 
 
\subsection{B. Magnetic field along the azimuthal direction}

\renewcommand{\theequation}{B-\arabic{equation}}
\setcounter{equation}{0}  
\renewcommand{\thefigure}{B-\arabic{figure}}
\setcounter{figure}{0}  

\begin{figure*}[htb]
 \centering
 \includegraphics[width=5.5 in]{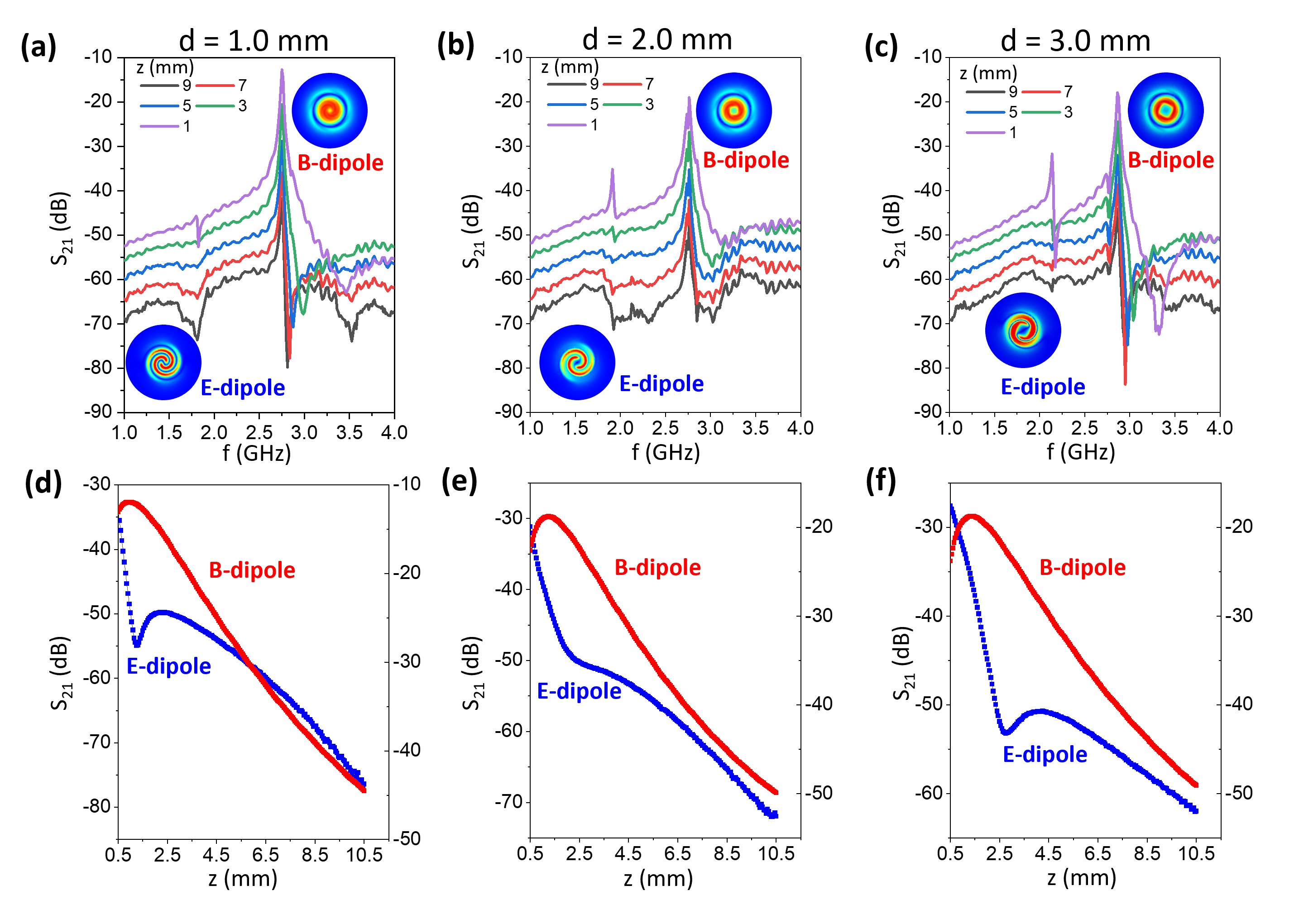}
 \caption{The measured transmission through a loop probe atop the resonator board at the location of the center disc edge (maximal coupling) and at selective probe height, $z=$ 1, 3, 5, 7, 9 mm, for (a) 4-S ($d=1.0$ mm), (b) 4-M ($d=2.0$ mm), (c) 4-L ($d=3.0$ mm). The corresponding azimuthal($z$)-dependence of the $B_z$-field for the fundamental E- and B-dipole modes for the (d) 4-S, (e) 4-M, (f) 4-L resonators. }
 \label{fig:z_map}
\end{figure*}

\renewcommand{\theequation}{C-\arabic{equation}}
\setcounter{equation}{0}  
\renewcommand{\thefigure}{C-\arabic{figure}}
\setcounter{figure}{0}  

\begin{figure*}[htb]
 \centering
 \includegraphics[width=7.0 in]{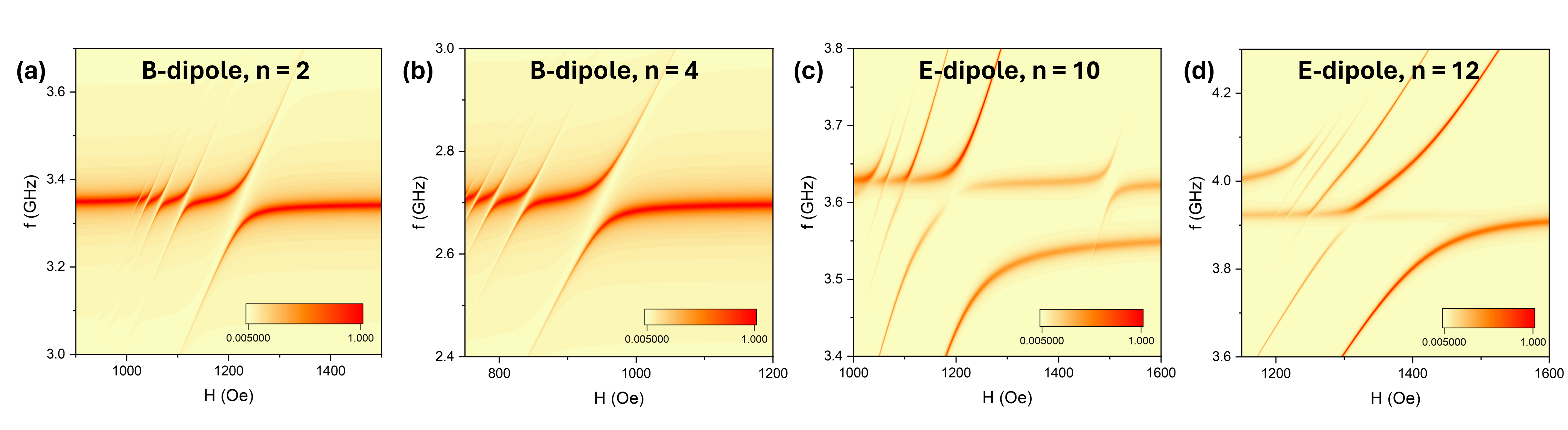}
 \caption{Additional modeling of photon-magnon coupling using coupled harmonic oscillator model for (a) 2nd order B-dipole, $n$ = 2, (b) B-dipole, $n$ = 4, (c) E-dipole, $n$ = 10, (d) E-dipole, $n$ = 12.}
 \label{fig:append_model}
\end{figure*}

Likewise, we can study the azimuthal dependence of the magnetic field using the fine-$z$ stage to control the probe position atop the board surface, as illustrated in Fig.\ref{fig:map_setup}. We calibrate our loop probe at a position much farther away from the board where the interaction is almost absent ($S_{21} < -90$ dB) and scan back to the board surface (at $\sim$ 0.5 mm). As an example, we studied the 4-S, 4-M, and 4-L series and focused on the fundamental E- and B-dipole modes. 

Figure \ref{fig:z_map}(a-c) shows the $S_{21}$ spectrum measured at selective height atop the board surface for the 4-S, 4-M, and 4-L boards, respectively. The B-dipole modes are generally stronger than the E-dipole modes in all three cases, which is in agreement with their overall mode profiles. 

Figure \ref{fig:z_map}(d-f) plots the $z$-dependent amplitude for the E- and B-dipole modes for the three cases. As a general observation, the E-dipole mode initially decays rapidly near the board surface, followed then by a much slower decay, after passing a small cusp around $\sim$ 2 - 3 mm. The initial decay relates to the center aperture, and thus is more rapid as the center aperture becomes smaller; this is due to the fact that the center aperture is the location where the magnetic fields from all the arms confluence. On the other hand, the slower decay at the 2nd stage pertains to the total magnetic field including the spiral arm regime. 

In addition, the position of the amplitude cusp is higher as the center aperture is smaller; this feature qualitatively reflects the `portion' of the total magnetic field that are due to the center aperture contribution: for smaller center aperture ($d$), such a portion is also smaller given the same spiral pattern diameter ($D$).  

The B-dipole mode, on the other hand, exhibits a different qualitative character. The mode amplitude first increases a bit with the height distance and peaks around $\sim$ 1 - 2 mm atop the board surface, which is then followed by a monotonic decay away from the surface. The peak amplitude is inversely proportional to the aperture size, and the peak location increases as the aperture size increases.

\subsection{C. Modeling of photon-magnon coupling}

The equations of motion for two coupled harmonic oscillators \cite{tay2018eit}, with one being driven by a plunger with constant frequency $\omega$, can be written as: 
\begin{equation} \label{eq:O1}
\ddot{x_1} + \omega_1^2 x_1 + 2\Gamma_1\omega_1\dot{x_1} - g_{12}x_2 = fe^{-i\omega t} 
\end{equation}
\begin{equation} \label{eq:O2}
\ddot{x_2} + \omega_2^2 x_2 + 2\Gamma_2\omega_2\dot{x_2} - g_{12}x_1 = 0 
\end{equation}

Where $\omega_1$ and $\omega_2$ are the resonant frequencies of the resonators, $\Gamma_1$ and $\Gamma_2$ are their damping coefficients, and \textit{g} is the coupling strength between them. \(x_1\) and \(x_2\) can be expressed as plane waves \((x_1, x_2)\) = \((A_1, A_2)e^{-i\omega t}\). Using this representation, along with a normalization, Equations \ref{eq:O1} and \ref{eq:O2} can be written in matrix form as:  
\begin{equation} \label{eq:matrix1}
\begin{pmatrix} 
\omega^2 - \omega_1^2 + 2i\Gamma_1\omega_1\omega & g_{12}\\ g_{12} & \omega^2 - \omega_2^2 + 2i\Gamma_2\omega_2\omega 
\end{pmatrix}
\begin{pmatrix}
A_1\\A_2
\end{pmatrix}
= \begin{pmatrix}
1\\0
\end{pmatrix}
\end{equation}

A rotating wave approximation \cite{wu2007twolevel} near the crossing point can be made, where the detuning \(\Delta\omega<< \omega_c + \omega_m\), along with the substitution $\tilde{\omega}_n = \omega_n - i\Gamma_n$, which represents the resonance frequency as a Lorentzian with center $\omega_n$ and width $\Gamma_n$. Eq. \ref{eq:matrix1} can now be written as:

\begin{equation}
\begin{pmatrix} 
\omega - \tilde{\omega}_1 & g_{12}\\ g_{12} & \omega - \tilde{\omega}_2 
\end{pmatrix}
\begin{pmatrix}
A_1\\A_2
\end{pmatrix}
= \begin{pmatrix}
1\\0
\end{pmatrix}
\end{equation}

For systems such as the ones modeled in this paper, where there are many modes interactions, the harmonic oscillator EOM can be generalized to include up to $n(n - 1)$ interactions, where \textit{n} is the total number of interacting quasiparticles. 

\begin{equation}
\ddot{x_1} + \omega_1^2 x_1 + 2\Gamma_1\omega_1\dot{x_1} - \sum_{k = 1}^{n} g_{1k} x_k = fe^{-i\omega t}
\end{equation}

An analogous matrix treatment of the system of \textit{n} quasiparticle EOMs generates the coupling matrix shown in Eq. \ref{eq:discmatrix}.

\renewcommand{\theequation}{C-\arabic{equation}}
\setcounter{equation}{0}  
\renewcommand{\thefigure}{C-\arabic{figure}}
\setcounter{figure}{0}  

The remainder of the resonator models are shown in Fig.\ref{fig:append_model}. For $n$ = 10 and $n$ = 12, the $n_z$ = 2 modes split into two, with the coupling between each mode and the associated magnon modes being unequal. Such a feature was observed in both the experiment and the simulation. The mode splitting greatly complicates the spectra, but the model being able to faithfully reproduce the experimental data is a consequence of the linear harmonic coupling between each photon and magnon mode. 

\bibliography{sample}

\end{document}